\documentstyle[psfig,twocolumn,prb,aps]{revtex}

\begin{document}
\draft

\twocolumn[\hsize\textwidth\columnwidth\hsize\csname %
 @twocolumnfalse\endcsname

\title{Finitely Correlated Generalized Spin Ladders}

\author{A. K. Kolezhuk\protect\cite{perm}}
\address{Institut f\"{u}r Theoretische Physik,
Universit\"{a}t Hannover, Appelstr. 2, 30167 Hannover, Germany\\
Institute of Magnetism, National Academy of 
Sciences and Ministry of Education of Ukraine\\
36(b) Vernadskii avenue, 252142 Kiev, Ukraine}

\author{H.-J. Mikeska\protect\cite{email}}
\address{Institut f\"{u}r Theoretische Physik,
Universit\"{a}t Hannover, Appelstr. 2, 30167 Hannover, Germany}

\date{March 13, 1998; printed out on \today }

\maketitle

\begin{abstract}
We study two-leg $ S={1\over2}$ ladders with general isotropic
exchange interactions between spins on neighboring rungs, whose ground
state can be found exactly in a form of finitely correlated (matrix
product) wave function.  Two families of models admitting an exact
solution are found: one yields translationally invariant ground states
and the other
describes spontaneously dimerized models with twofold degenerate
ground state.  Several known models with exact ground states
(Majumdar-Ghosh and Shastry-Sutherland spin-${1\over2}$ chains,
Affleck-Kennedy-Lieb-Tasaki spin-$1$ chain, $\Delta$-chain, Bose-Gayen
ladder model) can be obtained as  particular cases from the general
solution of the first family, which includes also a set of models with
only bilinear interactions.  Those two
families of models have nonzero intersection, which enables us to
determine exactly the phase boundary of the second-order transition
into the dimerized phase and to study the properties of this
transition.  The structure of elementary excitations in the dimerized
phase is discussed on the basis of a variational ansatz. For a
particular class of models, we present {\em exact\/} wave functions of
the elementary excitations becoming gapless at second-order transition
lines. We also propose a generalization of the Bose-Gayen model which
has a rich phase diagram with all phase boundaries being exact.

\end{abstract}

\pacs{75.10.Jm, 75.30.Kz, 75.40.Cx, 75.40.Gb}

]

\section{Introduction}
\label{sec:intro}

In recent years, Heisenberg spin ladders have attracted considerable
attention, particularly motivated by their peculiar properties of
being intermediate systems between dimensions one and two, as well
as by the hope to get some insight into the physics of metal-oxide
superconductors. \cite{DagottoRice96} It is now well established that
``regular'' (i.e., with only longitudinal ``leg'' and transverse
``rung'' exchange  couplings)
$S={1\over2}$ isotropic spin ladders with even number of legs have a
spin liquid ground state with short-range correlations and an
energy gap, while odd-legged ladders have a quasi-long-range ordered
gapless ground state. 

On the other hand, ``generalized'' ladders including other couplings
beyond the simplest case of rung and leg exchange are interesting toy
models which can interpolate between a variety of systems, exhibiting
remarkably rich behavior.
\cite{BoseGayen93+,BMN96,BKMN98,KM97,NersesyanTsvelik97,KM98prl,Weihong+}
Diagonal
interactions cause frustration and can change the structure of the
ground state;\cite{BoseGayen93+,Weihong+} recently it has been also
shown that biquadratic interactions are important
since they tend to produce dimerization and may lead to phase
transition into a ``non-Haldane'' spin liquid state with absence of
magnon excitations. \cite{NersesyanTsvelik97,KM98prl} In real magnetic
systems biquadratic terms can arise due to effective spin-spin
interaction mediated by phonons. \cite{NersesyanTsvelik97}

In the present paper we study the class of generalized $S={1\over2}$ ladder
models which admit an exact solution for the ground state in terms of the
so-called finitely correlated, or matrix product (MP)
states.\cite{Fannes+,Klumper+} The technique of MP states appears as a
natural generalization of the well known valence bond solid (VBS)
states,\cite{AKLT} which allows one to construct rather complicated
wave functions with given quantum numbers\cite{KMY97} and handle them
easily.  For standard Heisenberg spin models those MP states can serve
as good trial wave functions, \cite{BMN96,KMY97} while for certain
``perturbations'' of standard models those wave functions become exact
ground states.\cite{Klumper+,KM97,KM98prl} MP states have also proved
to be a convenient tool for variational study of elementary
excitations. \cite{NM96,TotsukaSuzuki95,BKMN98}

We use two MP ans\"atze for the ground state wave function having
different translational symmetry properties: one is invariant under
translation for one ladder rung, and for the other ansatz the
elementary cell consists of two rungs. Respectively, we obtain two
families of Hamiltonians with exact ground states; one should
emphasize that the Hamiltonian itself in both cases is translationally
invariant, so that the second family describes models with spontaneous
dimerization.  (Those families were studied by us recently
\cite{KM97,KM98prl} in a somewhat more restrictive formulation.)  The
two families have non-empty intersection, which makes possible to
determine exactly the boundary of a second-order phase transition from
translationally invariant to dimerized phase.
 
Thus we show that, in contrast to the common belief,
\cite{OstlundRommer95} {\em finitely correlated states can describe
critical models.\/} However, it turns out that in the dimerized phase
the dimer order parameter $D_{i}={\mathbf S}_{1,i}\cdot({\mathbf
S}_{1,i+1}-{\mathbf S}_{1,i-1})$ and spin correlation functions behave
in a rather peculiar way at the transition point $\tau=\tau_{c}$
(where $\tau$ is certain model parameter): (i) the spin correlation
function $\langle S_{i}^{z} S_{i+n}^{z}\rangle =A_{S}e^{-n/\xi}$
always decays exponentially, and the correlation length $\xi$ either
does not exhibit any singularities, or diverges as
$(\tau-\tau_{c})^{-2}$, but in the latter case the prefactor
$A_{S}\propto (\tau-\tau_{c})$ vanishes exactly at the transition,
thus preventing the appearance of the long-range spin order; (ii) the
dimer order correlation function $\langle D_{i}D_{i+n}\rangle \propto
(\tau-\tau_{c})^{2}$ does not depend on the distance $n$ and is just a
constant smoothly vanishing at the transition.

For excitations in the spontaneously dimerized phase we propose a
simple variational MP-type ansatz describing the elementary excitation
as a pair of solitons in dimer order; we show that the variational
estimate for the gap goes to zero at the phase boundary.

We show that for the certain class of translationally invariant models
having effectively a spin-$1$ Affleck-Kennedy-Lieb-Tasaki (AKLT)
\cite{AKLT} VBS ground state (formed by only triplet degrees of
freedom at each rung) it is possible to write an MP-type wave function
for the singlet excitation which is an exact
eigenstate of the Hamiltonian and whose gap closes at two different
second-order phase transition boundaries. For another class of models
whose ground state is a simple product of singlet bonds along the
ladder rungs, we give the condition of existence of the exact triplet
excitation. 

We also consider in more detail a toy model being a matrix-product
solvable ``perturbation'' of the ``composite spin'' model first
introduced by Bose and Gayen \cite{BoseGayen93+} and generalized later
by Weihong {\em et al.,\/} \cite{Weihong+} in the same sense as the
AKLT model is a perturbation of the standard $S=1$ Heisenberg spin
chain. This model is remarkable because one can obtain exactly its
full phase diagram, including several lines of phase transitions of
the first and second order.

The paper is organized as follows: in Sect.\ \ref{sec:routine} we
explain the general procedure of finding a set of exact MP-type
solutions, and sections \ref{sec:trans-inv} and \ref{sec:dim} describe
the two families of solutions mentioned above; in Sect.\ \ref{sec:dim}
we also discuss the boundaries separating dimerized and non-dimerized
phases and the structure of elementary excitations in the dimerized
phase. Section \ref{sec:BG+} is devoted to the generalized Bose-Gayen
model mentioned above and, finally, Sect. \ref{sec:summary} gives a
brief summary.

\section{Finitely correlated exact ground states: the construction routine}
\label{sec:routine}

\begin{figure}
 \mbox{\psfig{figure=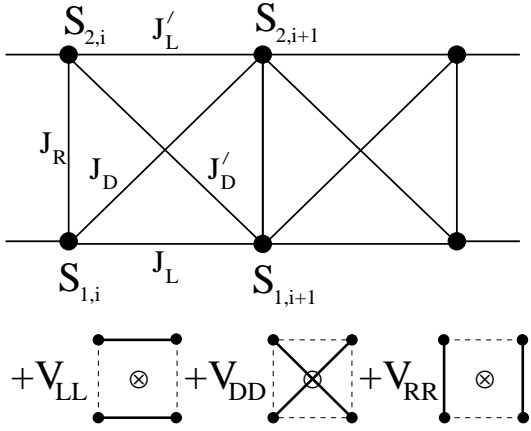,width=70mm,angle=-90}}
 \vspace{3mm}
\caption{\label{fig:genlad}
Generalized $S={1\over2}$ spin ladder described by the Hamiltonian
(\protect\ref{ham}), $V$'s denote the biquadratic couplings. 
}
\end{figure}

We start from a general form of the isotropic translationally
invariant spin ladder Hamiltonian with exchange interaction only
between spins on plaquettes formed by neighboring rungs, $\widehat H =
\sum_i \widehat h_{i,i+1}$, where the local Hamiltonian $\widehat{h}$
is defined as
\begin{eqnarray} 
\label{ham} 
\widehat h_{i,i+1}&=&
 {1\over2}(J_R+\varepsilon_{R}) {\mathbf S}_{1,i}{\cdot\mathbf S}_{2,i}
  + {1\over2}(J_R-\varepsilon_{R}) {\mathbf S}_{1,i+1}{\cdot\mathbf
 S}_{2,i+1} \nonumber \\
&+&J_L {\mathbf S}_{1,i}{\cdot\mathbf S}_{1,i+1} 
 +J_L' {\mathbf S}_{2,i}{\cdot\mathbf S}_{2,i+1}
+ J_D {\mathbf S}_{1,i} {\cdot\mathbf S}_{2,i+1}\nonumber\\
 &+&J_D' {\mathbf S}_{2,i}{\cdot\mathbf S}_{1, i+1}
+V_{LL} ({\mathbf S}_{1,i}{\cdot\mathbf S}_{1,i+1}) 
             ({\mathbf S}_{2,i}{\cdot\mathbf S}_{2,i+1})\nonumber\\
&+&  V_{DD} ({\mathbf S}_{1,i}{\cdot\mathbf S}_{2,i+1})  
               ({\mathbf S}_{2,i}{\cdot\mathbf S}_{1,i+1}) \\
&+& V_{RR} ({\mathbf S}_{1,i}{\cdot\mathbf S}_{2,i})
          ({\mathbf S}_{1,i+1}{\cdot\mathbf S}_{2,i+1}) -E_0 \,,\nonumber 
\end{eqnarray}
here the indices $1$ and $2$ distinguish lower and upper legs, and $i$
labels rungs. The model is schematically represented in Fig.\
\ref{fig:genlad} For periodic boundary conditions $J_R$ is the
coupling on the rungs, and Hamiltonians with different
$\varepsilon_{R}$ are physically indistinguishable; however, we
will see later that it is necessary to introduce $\varepsilon_{R}$ in
order to include models with inequivalent legs. In addition,
there are generally four different bilinear exchange couplings on the
legs and diagonals and three biquadratic terms. Together with the
constant term $-E_0$ we have a total of ten parameters.
For later convenience, we will also use  the
following notation:
\[
J_{L,D}^{(\pm)}={1\over2}(J_{L,D}\pm J_{L,D}')\,.
\]

We look for the ground state wave function $\Psi_{0}$ in a form of a
so-called {\em finitely correlated,\/} \cite{Fannes+} or {\em matrix
product (MP)\/} \cite{Klumper+}
state, and we use the following two ans\"atze for $\Psi_{0}$:
\begin{eqnarray}
\label{gs}
&& \Psi_0^{\text{inv}}  = \prod_i  g_i(u),\quad
\Psi_{0}^{\text{dim}}= \prod_{n} g_{2n-1}(u_{1})\cdot g_{2n}(u_{2}),\\
&& g_i(u)  =  \left[ 
\begin{array}{lr}
u | s \rangle_i + | t_0\rangle_i & - \sqrt{2}  | t_{+1} \rangle_i \\
\sqrt{2}  | t_{-1} \rangle_i & u | s \rangle_i -  | t_0 \rangle_i
\end{array} 
\right].
\end{eqnarray}
Here $|s\rangle_{i}$ and $|t_{\mu}\rangle_{i}$ are the singlet and
triplet states of the $i$-th rung, and $u_{1}$, $u_{2}$ are free
parameters. Strictly speaking, $\Psi_{0}$ in (\ref{gs}) is a matrix
whose trace should be taken to get the wave function of a ladder with
periodic boundary conditions. Three other linear independent wave
functions $\mbox{tr}(\sigma^{\mu}\Psi_{0})$, $\mu=0,\pm1$ describe
states of a ladder with open ends having different behavior at the
edges, which are similar to the edge states in the effective $S=1$
chain. \cite{BMN96,KMY97} The ansatz (\ref{gs}) obeys rotational
symmetry, i.e., $\mbox{tr}(\Psi_{0})$ is a global singlet, and the
states $\mbox{tr}(\sigma^{\mu}\Psi_{0})$ form a spin-$1$ triplet.  The
state $\Psi_{0}^{{\text{inv}}}$ is translationally invariant under the
translation for one ladder rung, and $\Psi_{0}^{\text{dim}}$ describes
dimerized states with spontaneously broken translational invariance
(unless $u_{1}=u_{2}$ when it obviously becomes identical to
$\Psi_{0}^{\text{inv}}$). Translation of $\Psi_{0}^{\text{dim}}$ for
one rung leads to a different state with the same energy.  The
construction (\ref{gs}) was originally proposed as a variational wave
function; \cite{BMN96} later the translationally invariant form
$\Psi_{0}^{\text{inv}}$ was used by us \cite{KM97} to construct a
class of exact ground states for the model (\ref{ham}) under somewhat
restrictive assumptions $\varepsilon_{R}=0$, $J_{L}=J_{L}'$,
$V_{LL}=V_{DD}$, $V_{RR}=0$, and we have reported several particular
solutions of the type $\Psi_{0}^{\text{dim}}$ in Ref.\
\onlinecite{KM98prl}.

One can observe \cite{BMN96} that ans\"atze (\ref{gs})  describe
several known examples of VBS-type states, e.g.,
$\Psi_{0}^{\text{inv}}$ at $u=0$  yields
the ground state of the effective AKLT chain \cite{AKLT} whose $S=1$
spins are composed from pairs of $S={1\over2}$ spins of the ladder
rungs,
\begin{equation}
\label{S1biquad}
\widehat{H}=\sum_{n} {\mathbf
S}_{n} {\mathbf S}_{n+1}
+{1\over3} ({\mathbf S}_{n}\cdot {\mathbf S}_{n+1})^{2},
\end{equation}
while for $u=1$ and $u=\infty$ one obtains
two degenerate dimer ground states of the Majumdar-Ghosh (MG) model,
\cite{MajumdarGhosh69} with singlets residing on the diagonals and on
the rungs, respectively [the MG model results from (\ref{ham}) at
$V_{RR,LL,DD}=0$, $J_{D}'=0$, $J_{L}=J_{L}'$,
$J_{R}=J_{D}=2J_{L}$]. The wave function $\Psi_{0}^{\text{dim}}$ at
$u_{1}=-u_{2}=\pm 1$ describes a state with checkerboard-type
ordered singlet bonds along the ladder legs. Thus one may think of the
construction (\ref{gs}) as of interpolating between several types of
VBS states mentioned above.
 
The construction of exact ground states can be performed in the way
 outlined in Refs.\ \onlinecite{Klumper+,NiggZitt96}: one has to require
 $\Psi_{0}$ to be a zero-energy ground state of the local Hamiltonian
 $\widehat{h}_{i,i+1}$, which ensures that it is a ground state of the
 global Hamiltonian $\widehat{H}$. This yields the following
 conditions:

(i) for $\Psi_{0}^{\text{inv}}$, the local Hamiltonian
$\widehat{h}_{12}$ should annihilate all states being matrix elements
of the product $g_{1}(u) \cdot g_{2}(u)$:
\begin{equation} 
\widehat{h}_{12}
\big\{g_{1}(u) \cdot g_{2}(u)\big\}=0\,,
\end{equation}
while
for $\Psi_{0}^{\text{dim}}$ it is necessary that
$\widehat{h}_{12}$ annihilates all states contained in 
the two products $g_{1}(u_{1})\cdot g_{2}(u_{2})$ and
$g_{1}(u_{2})\cdot g_{2}(u_{1})$, 
\begin{mathletters}
\label{cond1} 
\begin{eqnarray} 
&&\widehat{h}_{12} \big\{
g_{1}(u_{1}) \cdot g_{2}(u_{2})\big\}=0\,,\nonumber\\
&&\widehat{h}_{12}
\big\{g_{1}(u_{2}) \cdot g_{2}(u_{1})\big\}=0\,;
\end{eqnarray}
\end{mathletters}

(ii) all the other eigenstates of $\widehat{h}_{12}$ have positive
energy. Then $\Psi_{0}$ is the zero-energy ground state of
$\widehat{H}$; if one drops the constant term $-E_{0}$ in (\ref{ham}),
the remaining Hamiltonian has the energy density $E_{0}$ per rung.

It is convenient to write the local Hamiltonian $\widehat{h}_{i,i+1}$
in terms of projectors on the states with fixed angular momentum
$|\psi_{j\mu}^{(k)}\rangle$ of the two-rung plaquette $(i,i+1)$.  The
complete set of the plaquette states contains one quintuplet ($j=2$),
three triplets ($j=1$) and two singlets ($j=0$), therefore a general
form of $\widehat{h}$ reads as:
\begin{eqnarray} 
\label{ham-proj}
\widehat h_{i,i+1} 
&=& \lambda_{2} \sum_\mu  |\psi_{2\mu}\rangle \langle \psi_{2\mu} |
  + \sum_{k,l=1}^{3} \lambda_1^{(k,l)} 
\sum_\mu |\psi^{(k)}_{1\mu} \rangle \langle\psi^{(l)}_{1\mu} | \nonumber\\
     &+& \sum_{k,l=1,2} \lambda_0^{(k,l)} 
     |\psi^{(k)}_{00} \rangle \langle\psi^{(l)}_{00}|
\end{eqnarray}
Here obviously $\lambda_{J}^{(k,l)}=(\lambda_{J}^{(l,k)})^{*}$ because
of the hermitian property of $\widehat{h}$; ten independent 
$\lambda$'s are linearly related to the ten
constants in the Hamiltonian of Eq.\ (\ref{ham}).

For each total momentum $j$, the complete set of the plaquette states
can be divided into two subsets: $n_{j}^{g}$ {\em local ground
states\/} $|\psi_{j\mu}^{(g,k)}\rangle$ which are contained in the
matrix products $g_{i}g_{i+1}$, and $n_{j}^{e}$ {\em local
eigenstates\/} $|\psi_{j\mu}^{(e,k)}\rangle$ which do not enter there;
respectively the set of indices $\{k\}$ for each $j$ can be divided in
two subsets which we will denote ${\cal G}_{j}$ for local ground
states and ${\cal E}_{j}$ for local eigenstates.

The conditions (i) mean that the local Hamiltonian $\widehat{h}$
should project only onto the states $|\psi_{j\mu}^{(e,k)}\rangle$, and
the multiplets $|\psi_{j\mu}^{(g,k)}\rangle$ have to be absent in Eq.\
(\ref{ham-proj}). This results in the following equations:
\begin{equation} 
\label{zeroeqs} 
\lambda_{j}^{(k,l)}=0,\quad l\in {\cal G}_{j},\; \forall k
\end{equation}
being essentially a system of linear equations in the Hamiltonian
coupling constants $J_{..}$, $\varepsilon_{..}$, $V_{..}$, $E_{0}$
[see Eq.\ (\ref{ham})]. The only nonzero $\lambda$'s are then
$\lambda_{j}^{(k,l)}$ for both $k$ and $l$ belonging to ${\cal
E}_{j}$, 
so that hereafter we
denote them simply as $\lambda_{j}^{(p,p')}$, $p,p'=1,\ldots,n_{j}^{e}$.

The conditions (ii) require that all the eigenstates within the
subspace determined by the set of states $|\psi_{j\mu}^{(e,k)}\rangle$
have positive energy, which yields the following inequalities:
\begin{equation} 
\label{lambdaeqs}
\widetilde{\lambda}_{j}^{(\alpha)} \geq 0\,,\quad 
\alpha=1\ldots n_{j}^{e}\,, 
\end{equation}
where $\widetilde{\lambda}_{j}^{(\alpha)}$ denotes the $\alpha$-th
eigenvalue of the matrix
\[
\Lambda_{j}=[\lambda_{j}^{(p,p')}],\quad p,p'=1,\ldots,n_{j}^{e}
\]
If one or more of $\widetilde{\lambda}_{j}^{(\alpha)}$ is zero, this
indicates a possibility for an additional degeneracy of the ground
state (quantum phase transition).
However, generally the conditions of stability imposed by the
inequalities (\ref{lambdaeqs}) are only {\em sufficient\/} but not
{\em necessary,\/} and in order to conclude on the presence of a phase
transition, one should be able to construct the other ground state
explicitly. Later we will see that in many important cases this is
indeed possible.

\section{Models with unbroken translational symmetry}
\label{sec:trans-inv}

For $\Psi_{0}^{\text{inv}}$, it is easy to check that generally the matrix
product $g_{1}(u)\cdot g_{2}(u)$ contains only the following two
multiplets which should become local ground states:
\begin{eqnarray}
\label{lgs-inv}
&&(3 + u^4)^{-1/2} \left( u^2 | ss \rangle
-   \sqrt{3} | tt \rangle_{00} \right) \equiv
| \psi_{00}^{(g)} \rangle \,,\\
&&(1 + u^2)^{-1/2}
\left \{ {u\over\sqrt{2}} ( | t_{\mu}s \rangle + | st_{\mu} \rangle ) - 
| tt \rangle_{1\mu} \right \} \equiv | \psi_{1\mu}^{(g)} \rangle \,;\nonumber
\end{eqnarray}
(in fact, the second multiplet above enters
only in the combination $(3+u^{2})^{-1}| \psi_{1\mu}^{(g)} \rangle$,
so that $| \psi_{1\mu}^{(g)} \rangle$ disappears from $g_{1}(u)\cdot
g_{2}(u)$ in the limit $u\to\infty$, see below).  The remaining
multiplets becoming local eigenstates can be chosen as \cite{KM97}
\begin{eqnarray}
\label{les-inv}
&& | \psi_{00}^{(e)} \rangle =  (3  +u^4)^{-1/2} 
\left(  \sqrt{3} | ss \rangle
+ u^2 | tt \rangle_{00} \right) \nonumber \\
&& | \psi_{1\mu}^{(e,1)} \rangle = {1\over\sqrt{2}} \big(| t_{\mu}s \rangle -
| st_{\mu} \rangle \big) \\
&& | \psi_{1\mu}^{(e,2)} \rangle =
(1+u^2)^{-1/2} 
 \left\{ {1\over\sqrt{2}} 
\big(| t_{\mu}s \rangle + | st_{\mu} \rangle \big) + 
  u | tt \rangle_{1\mu} \right\} \nonumber \\
&& | \psi_{2\mu}^{(e)} \rangle = | tt \rangle_{2\mu}\,.\nonumber
\end{eqnarray}
Here $|tt\rangle_{1\mu}$  and $|tt\rangle_{00}$ denote respectively
the triplet ($j=1$) and the singlet ($j=0$) constructed from two
triplet states on rungs $i$ and $i+1$. 

Equations (\ref{zeroeqs}) [in this case
$n_{0}^{g}=n_{0}^{e}=1$, $n_{1}^{g}=1$, $n_{1}^{e}=2$, $n_{2}^{e}=1$,
$n_{2}^{g}=0$], which impose the demand that the local Hamiltonian
annihilates all the states (\ref{lgs-inv}), can be rewritten in a form
of relations between the Hamiltonian parameters as follows:
\begin{eqnarray} 
\label{gensol-inv} 
&&J_{R} = {5\over6}\lambda_{2}  +{AB\over6}\lambda _{0}  
-{1\over2}\big(\lambda_{1}^{(1,1)}-C \lambda _{1}^{(2,2)}\big), \nonumber\\
&&J_{L}^{(+)}={5\over12}\lambda_{2}-{A(A+B)\over12}\lambda_{0}
-{1\over4}\big(\lambda_{1}^{(1,1)}
+C\lambda_{1}^{(2,2)}\big),\nonumber\\
&&J_{D}^{(+)}={5\over12}\lambda_{2}-{B(A+B)\over12}\lambda_{0}+{1\over4}
\big(\lambda_{1}^{(1,1)}-\lambda_{1}^{(2,2)}\big),\nonumber\\
&&V_{RR}={1\over3}\lambda_{2}-{AB\over3}\lambda_{0}
-\lambda_{1}^{(1,1)}+C\lambda_{1}^{(2,2)}, \\
&&V_{LL}={1\over3}\lambda_{2}+{A(A+B)\over3}\lambda_{0}
-\lambda_{1}^{(1,1)}-C\lambda_{1}^{(2,2)}, \nonumber\\
&&V_{DD}={1\over3}\lambda_{2}+{B(A+B)\over3}\lambda_{0} 
+\lambda_{1}^{(1,1)}-\lambda_{1}^{(2,2)},\nonumber\\
&&\varepsilon_{R}={2\lambda _{1}^{(1,2)}\over \sqrt{1+u^{2}}},\;\;
J_{L}^{(-)}=-{u\lambda_{1}^{(1,2)}\over \sqrt{1+u^{2}}},\;\;
J_{D}^{(-)}=-{u\lambda_{1}^{(2,2)}\over 1+u^{2}},
\nonumber
\end{eqnarray}
the energy density per rung $E_{0}$ being
\begin{equation}
\label{e-inv}
E_{0}=-{5\over16}\lambda_{2}-{1\over16}\lambda_{0}
-{3\over16}\big(\lambda_{1}^{(1,1)}+\lambda_{1}^{(2,2)}\big)\,.
\end{equation}
Here we have used the following shorthand notation:
\[
 A={u^{2}+3\over (3+u^{4})^{1/2}},\quad
 B={u^{2}-3\over (3+u^{4})^{1/2}},\quad
C={u^{2}-1\over u^{2}+1},
\]
It is sufficient to consider only positive values of $u$, because
changing the sign $u\to-u$ amounts just to interchanging the ladder
legs and thus does not bring in any new physics. One can see that
the ``unphysical'' parameter $\varepsilon_{R}=-2J_{L}^{(-)}/u$ is just
needed to allow the ladder legs to be inequivalent.  The conditions
(\ref{lambdaeqs}) now take the form
\begin{eqnarray} 
\label{ineq-inv} 
&&\lambda_{0}\geq 0, \quad \lambda_{2}\geq 0,\quad
\lambda_{1}^{(1,1)}+\lambda_{1}^{(2,2)}\geq 0,\nonumber\\
&&\big[\lambda_{1}^{(1,2)}\big]^{2} 
\leq \lambda_{1}^{(1,1)} \lambda_{1}^{(2,2)}\,.
\end{eqnarray}

Equations (\ref{gensol-inv}) describe a five-parametric family of spin
ladder models (five $\lambda$'s and $u$, but one of the $\lambda$'s is
irrelevant since it just determines the energy scale) whose
translationally invariant ground state can be written exactly in the
form of a matrix product $\Psi_{0}^{\text{inv}}$ (\ref{gs}).

Though the limit $u\to\infty$ of Eqs.\ (\ref{gensol-inv}) is formally
well defined, the family (\ref{gensol-inv}) does not contain {\em
all\/} models having a product of singlet bonds on the rungs as their
exact ground states, e.g., the model of Bose and Gayen
\cite{BoseGayen93+} is missing there.  For $u=\infty$ obviously only
the state $|ss\rangle$ enters the matrix product $g_{1}\cdot g_{2}$,
and this case should be considered separately. The resulting general
solution for the $u=\infty$ case is
\begin{equation} 
\label{gensol-v0}
J_{L}^{(+)} -J_{D}^{(+)}=(V_{LL}-V_{DD})/4\,,
\end{equation}
the energy density per rung being
\begin{equation} 
\label{ev0} 
E_{0}=-3J_{R}/4+3(V_{LL}+V_{DD}+3 V_{RR})/16\,.
\end{equation}
The conditions of positivity (\ref{lambdaeqs}) result in the
constraints
\begin{eqnarray} 
\label{ineq-v0}
&&\mu_{0}=J_{R}-2J_{D}^{(+)}+(V_{DD}-V_{RR})/2
\geq 0,\nonumber\\ 
&&\mu_{2}=J_{R}+J_{D}^{(+)}
-(V_{DD}+2V_{RR})/4 \geq 0,\nonumber\\  
&&\widetilde{\mu}_{1}^{(i)}\geq 0,\quad i=1,2,3\,,
\end{eqnarray}
where $\widetilde{\mu}_{1}^{(i)}$ are the eigenvalues of a symmetric
3$\times$3 matrix $[\mu_{1}^{ij}]$, with the matrix elements defined as
\begin{eqnarray} 
\label{3x3v0}
&&\mu_{1}^{11}={J_{R}\over2} -{V_{LL}\over2} -{3V_{RR}\over4},\quad 
\mu_{1}^{22}={J_{R}\over2}-{V_{DD}\over2}-{3V_{RR}\over4},\nonumber\\
&&\mu_{1}^{12}=\varepsilon_{R}/2,\quad 
\mu_{1}^{13}=-J_{L}^{(-)}, \quad 
\mu_{1}^{23}=-J_{D}^{(-)}\nonumber\\
&&\mu_{1}^{33}=J_{R}-J_{L}^{(+)} 
-(2V_{DD}+2V_{RR}+V_{LL})/4\,.
\end{eqnarray}

One can check that the families (\ref{gensol-inv}) and
(\ref{gensol-v0}) match each other if $\lambda_{0}=0$, so that it is
natural to assume that vanishing $\lambda_{0}$ signals the first-order
phase transition into the rung-dimer phase. Indeed, imagine we have a
MP-solvable model with certain ground state and gradually decrease
$\lambda_{0}$ to $0$, then exactly at $\lambda_{0}=0$ the rung-dimer
state is degenerate with the ``old'' ground state, so that we have
level crossing. The same reasoning can be applied to the transition
into the completely polarized ferromagnetic phase which is driven by
vanishing $\lambda_{2}$; note that one cannot do the same for
$\widetilde{\lambda}_{1}^{(i)}$ because no uniform global wave
function can be constructed from the local Hamiltonian eigenstates
corresponding to those eigenvalues. 
However, later we will see that
simultaneous vanishing of $\lambda_{1}^{(1,1)}$ and
$\lambda_{1}^{(1,2)}$ can be associated with another phase transition,
the {\em second order\/} phase transition into a spontaneously
dimerized phase. In special cases, other eigenvalues may become
relevant for determining the phase transition points (see Sect.\ 
\ref{subsec:AKLT} and Sect.\ \ref{sec:BG+} below).

In the rest of this section we consider several particular solutions
belonging to the general families derived above.

\subsection{Models with only bilinear exchange coupling}

If one demands that all biquadratic interactions in (\ref{ham})
are absent, the general family of models (\ref{gensol-inv}) reduces to
the following two ``branches:''

{\em 1. Models with completely dimerized bonds,\/} whose ground state
is just a product of singlet dimers along the ladder diagonals, are
obtained from (\ref{gensol-inv}) by setting $u=1$, $\lambda_{1}^{(2,2)}=1$,
$\lambda_{1}^{(1,1)}=y$, $\lambda_{1}^{(1,2)}=x\sqrt{2}$, $\lambda_{0}=2y-1$,
$\lambda_{2}=2-y$:
\begin{eqnarray} 
\label{ss-delta} 
&&J_{R}=2(1-y),\quad J_{L}=1-y-x,\quad J_{L}'=1-y+x,\nonumber\\
&&J_{D}'=1,\quad J_{D}=0,\quad \varepsilon_{R}=2x\,,
\end{eqnarray}
the conditions (\ref{ineq-inv}) require that $1/2 \leq y \leq 2$,
$x^{2}\leq y/2$. It is easy to see that at $x=0$ this model describes
an $S={1\over2}$ zigzag chain with alternated nearest-neighbor (NN)
exchange and frustrating next-nearest neighbor (NNN) interaction, with
the alternation proportional to $2y-1$ and the ratio of NNN exchange
constant to the smaller NN one equal to ${1\over2}$. For
${1\over2}\leq y \leq 1$ the ground state of this model was found by
Shastry and Sutherland, \cite{SS81} and $y={1\over2}$ corresponds to
the well-known Majumdar-Ghosh (MG) model; \cite{MajumdarGhosh69} at
the MG point there is another, degenerate ground state with singlet
bonds on the rungs which is contained in the family (\ref{gensol-v0}).
At $x=y-1$ (\ref{ss-delta}) describes the so-called $\Delta$-chain, or
see-saw chain \cite{delta-chain} which, like the MG model, also has
two degenerate ground states at $y={1\over2}$.

The interval $1\leq y \leq 2$ corresponds to partially ferromagnetic
interactions, and at $y=2$ the model (\ref{ss-delta}) exhibits a
first-order phase transition into the completely polarized ferromagnetic
state.\cite{KM97} 

\begin{figure}
\mbox{\hspace{5mm}\psfig{figure=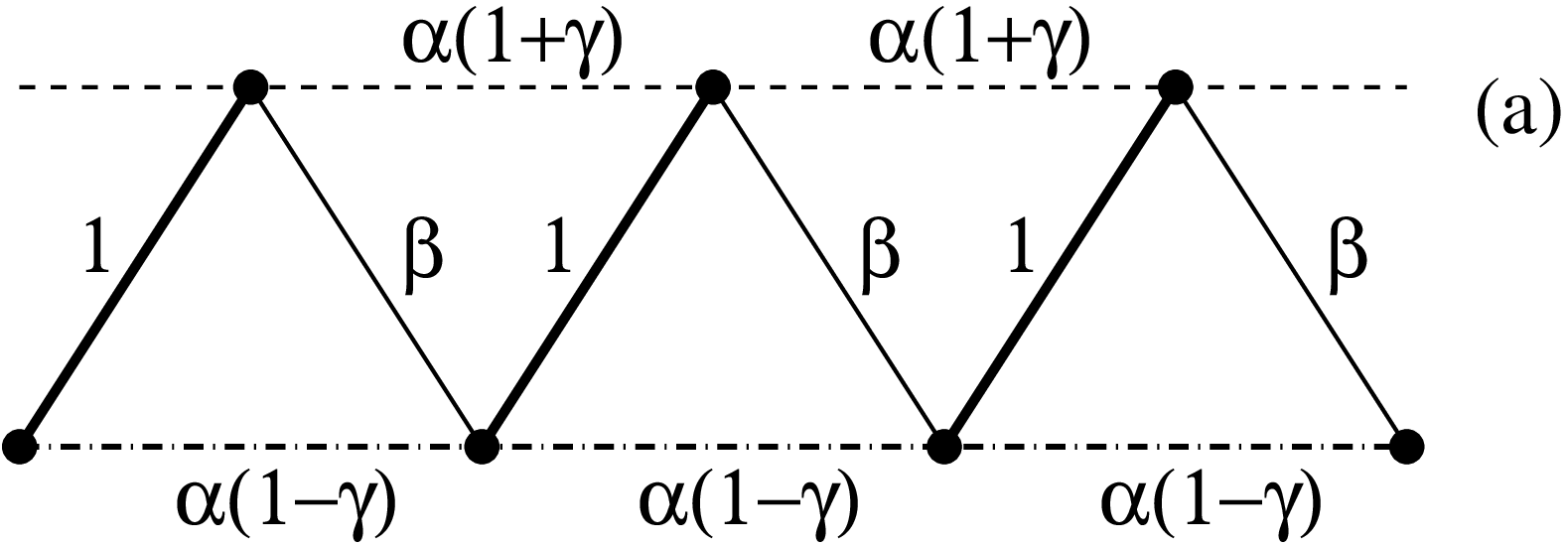,width=70mm,angle=-90}}
\vspace{3mm}
\mbox{\psfig{figure=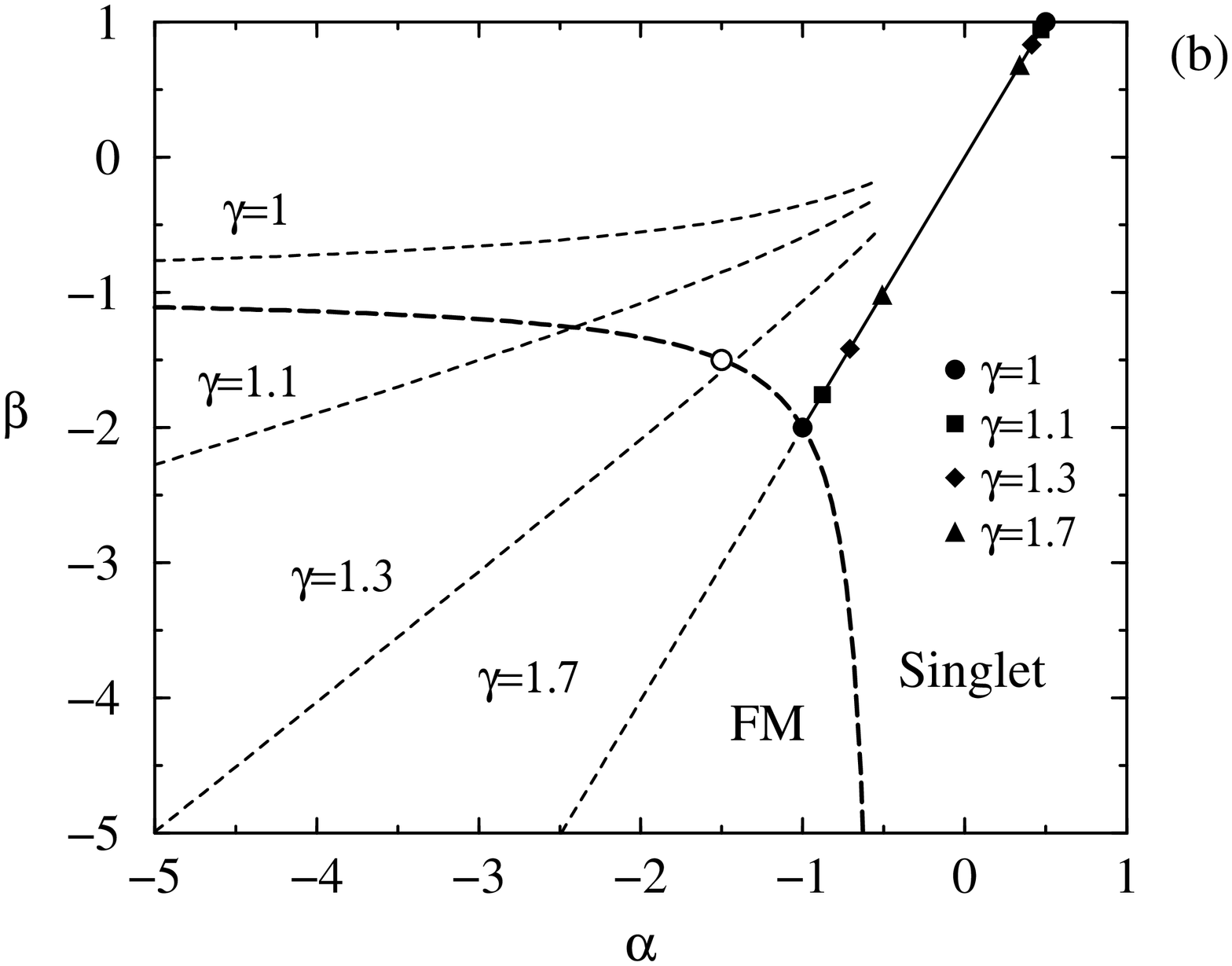,width=80mm,angle=-90}}
\vspace{3mm}
\caption{\label{fig:zigzag} (a) The generalized zigzag chain with two
unequal next-nearest neighbor couplings; (b) its phase diagram.  Thick
dashed line $\beta=F(\alpha)\equiv -2\alpha/(1+2\alpha)$ and thin
dashed lines $\beta=f_{\gamma}(\alpha)$ denote the boundaries of
stability of the ferromagnetic (FM) phase resulting from the first and
the second conditions in (\protect\ref{stabFM}), respectively; the
actual boundary between the FM phase and the singlet phase is
determined by the two conditions $\beta<F(\alpha)$, $\beta<
f_{\gamma}(\alpha)$.  The line $\beta=1$, $\alpha\leq \alpha_{c}\simeq
0.2411$ is gapless, the rest of the singlet phase is gapped.  For
$|\gamma|\leq 1$ the FM-singlet boundary is determined solely by the
thick dashed line, and for $|\gamma|>1$ the FM phase starts to shrink
with increasing $|\gamma|$ as indicated by the thin dashed lines.
On the thick solid line ($\beta=2\alpha$) the ground state is a
product of dimers along the $J=1$ bonds. For $|\gamma|\leq1$ it is stable
for $-1\leq \alpha \leq {1\over2}$, and $({1\over2},1)$ is the
generalized Majumdar-Ghosh point; for $|\gamma|>1$, according to the
criterion (\protect\ref{ssline}) obtained from the matrix product
calculation, the region of guaranteed stability of the dimer-product
ground state on the line $\beta=2\alpha$ shrinks with increasing
$|\gamma|$ as shown with the corresponding symbols (however, the MP
criterion generally gives the conditions which are only sufficient but
not necessary, so that the actual region of stability of the
dimer-product ground state may be wider, see the discussion in the
text). The point $(-{3\over2},-{3\over2})$ at the thick dashed line
(shown with an open circle) has a matrix-product state
$\Psi_{0}^{\text{inv}}$ of the type (\protect\ref{gs}) with $u=3$,
which is stable at least for $|\gamma|<1$.  }

\end{figure}

One can consider a more general model of a zigzag chain with two unequal
NNN interactions $\alpha(1\pm \gamma)$ and alternating NN couplings $1$ and
$\beta$, $\beta<1$, as shown in Fig.\ \ref{fig:zigzag}a. This model
interpolates between the usual frustrated chain ($\gamma=0$) and
the $\Delta$-chain ($\gamma=1$). For  fixed $\gamma$, the solution
(\ref{ss-delta}) determines the line
\begin{eqnarray} 
\label{ssline} 
&&\beta=2\alpha,\quad  \max(-1,\zeta_{-})\leq
\alpha\leq \min({1\over2},\zeta_{+}), \\
&&\zeta_{\pm}=\{ -1\pm(1+8\gamma^{2})^{1/2}\}/(4\gamma^{2})\,.\nonumber
\end{eqnarray}
in the $(\alpha\beta)$ plane where the ground state is a product of
NN dimers on $J=1$ bonds, see Fig.\ \ref{fig:zigzag}b. For
$|\gamma|\leq 1$ the minimal and maximal values of $\alpha$ in
(\ref{ssline}) are always $-1$ and ${1\over2}$, the point
$\alpha=-1$, $\beta=-2$  lies on the boundary of the transition
into the ferromagnetic (FM) phase, and $\alpha=1/2$, $\beta=1$ is the
(generalized) Majumdar-Ghosh point. As we will see below, the situation
for $|\gamma|>1$  is more complicated.

It is worthwhile to make an attempt to estimate the boundaries of
stability of the FM phase by using a standard spin wave theory.
 Introducing two kinds of Holstein-Primakoff bosons for
representing the spin operators on two different legs, and passing to
the momentum space, one obtains in the lowest order the following
spin wave Hamiltonian:
\begin{eqnarray*} 
&&\widehat{{\cal H}}=\sum_{k} \{ \varepsilon_{1k} a^{\dag}_{1k}a_{1k}
+\varepsilon_{2k} a^{\dag}_{2k}a_{2k} 
+(\Phi_{k}a^{\dag}_{1k}a_{2k}+\text{h.c.})\}\,,\\
&&\varepsilon_{1k,2k}=-{1+\beta\over2}-\alpha(1\pm \gamma)(1-\cos k),\;
\Phi_{k}={\beta+e^{-ik}\over2}\,.
\end{eqnarray*}
This Hamiltonian is trivially diagonalized, yielding the following
expressions for the dispersion of two magnon branches:
\begin{eqnarray} 
\label{dispFM} 
&&\varepsilon_{A,B}(k)=\varepsilon_{1k,2k}
+{(\varepsilon_{1k}-\varepsilon_{2k}) +2 |\Phi_{k}|
z_{k}\over(1+z_{k}^{2})}\,,
\\
&&z_{k}=(\varepsilon_{1k}-\varepsilon_{2k})/2|\Phi_{k}| 
+\sigma
\big\{ 1+(\varepsilon_{1k}-\varepsilon_{2k})^{2}/4|\Phi_{k}|^{2}
\big\}^{1/2}\,,\nonumber
\end{eqnarray}
here $\sigma\equiv\mbox{sgn}(\varepsilon_{1k}-\varepsilon_{2k})$. It
is easy to check that $\varepsilon_{A}(k)\simeq C k^{2}$ at $ k\to0$,
$C=-{1\over4}\{2\alpha+\beta/(1+\beta)\}$. Thus $\varepsilon_{A}$
represents the usual branch of ferromagnons, and the condition $C=0$
determines the point where this branch becomes unstable. The other
branch $\varepsilon_{B}$ represents ``optical'' ferromagnons,
$\varepsilon_{B}(k=0)=-1-\beta$. For $\gamma=0$ the branch
$\varepsilon_{B}$ is always gapped, but for large enough $|\gamma|$ it
closes at $k=\pi$. Thus the region of stability of the FM phase is
determined by the following two conditions:
\begin{equation} 
\label{stabFM}
\beta\leq -{2\alpha \over 1+2\alpha},\quad |\gamma| \leq
\gamma_{c}(\alpha,\beta)\,,
\end{equation}
where the critical point $\gamma_{c}$ is defined as a real positive
root of the following fourth-order polynomial equation:
\begin{eqnarray} 
\label{xcFM} 
64\alpha^{4} 
\gamma_{c}^{4}
&-&[32\alpha(2\alpha+\beta+1)+12\beta^{2}-8\beta+12]\alpha^{2} 
\gamma_{c}^{2}\nonumber\\
&-&(\beta-1)^{2}(1+2\alpha)(\beta+2\alpha)=0\,.
\end{eqnarray}

It is interesting to note that the curve in the phase space of the
model determined by the first condition in (\ref{stabFM}) at $\gamma=0$
gives precisely the set of models with exact singlet ground states 
degenerate with the FM state which were studied in Ref.\
\onlinecite{Dmitriev+97}. Thus at least the first condition in (\ref{stabFM}),
obtained from a simple spin wave calculation, remarkably coincides
with the exact result. It should be mentioned that the exact singlet
wave functions presented in Ref.\ \onlinecite{Dmitriev+97} exhibit
double spiral order, and are very different from our MP wave
functions, which indicates high degeneracy on the line of transition
from the FM to singlet phase.

The second condition in (\ref{stabFM}) becomes relevant
only at $|\gamma|>1$, i.e., when the leg couplings have different signs;
at $|\gamma|<1$ the boundary of FM phase is determined solely by the
first condition,
see Fig.\ \ref{fig:zigzag}b.

One can also observe that the critical value of $\gamma$ determined by
(\ref{xcFM}) is different from that which can be obtained from our MP
calculation; at $\alpha=-1$, $\beta=-2$ it follows from
(\ref{ssline}) that $|\gamma|\leq 1$, while from (\ref{xcFM}) one obtains
$|\gamma|<\gamma_{c}\simeq 1.7$. As we discussed before,
vanishing of any $\lambda$'s except $\lambda_{0}$ and $\lambda_{2}$
cannot be associated with a phase transition, and generally the
conditions (\ref{ineq-inv}), being sufficient but not necessary, tend
to {\em underestimate\/} the stability region.

{\em 2. Bilinear models with nontrivial ground states\/} result from
(\ref{gensol-inv}) when
\begin{eqnarray} 
&&\lambda_{2}=u^{2}(9-u^{2}), \quad
\lambda_{0}=2(3+u^{4}), \quad \lambda_{1}^{(1,1)}=5u^{2}+3, \nonumber\\
&&\lambda_{1}^{(2,2)}=(3+u^{2})(1+u^{2}),\quad \lambda_{1}^{(1,2)}=
x(1+u^{2})^{1/2}\,,
\label{2nd} 
\end{eqnarray}
which gives the Hamiltonian parameters as
\begin{eqnarray} 
\label{eqnarray}
&& J_{R}=6(u^{2}-1),\quad \varepsilon_{R}=2x,\nonumber\\
&& J_{L}^{(+)}=u^{2}(1-u^{2}),\quad
J_{L}^{(-)}=-ux,\nonumber\\ 
&& J_{D}=-u(3+u)(u-1)^{2},\quad J_{D}'=u(3-u)(u+1)^{2}\,,
\end{eqnarray}
with the energy density per rung $E_{0}=-{3\over2}(1+3u^{2})$.  From
(\ref{ineq-inv}) it follows that $x^{2}\leq (3+u^{2})(5u^{2}+3)$,
$|u|\leq 3$. At $u=3$ the eigenvalue $\lambda_{2}$ vanishes, i.e., the
system undergoes a phase transition into the completely polarized state,
and at the same time one of the diagonal interactions vanishes, making
the ladder equivalent to a frustrated chain; \cite{KM97} one can check
that this solution belongs to the transition line from FM to the singlet
phase determined by (\ref{stabFM}), and again the critical value of
$|\gamma|$ obtained from the MP calculation, $\gamma_{c}=1$, is
lower than one given by (\ref{xcFM}), $\gamma_{c}\simeq1.27$.

Another interesting limit is $u=1$,
when the rung coupling and one of the diagonal couplings are
simultaneously zero, and the ground state is a product of singlets
sitting on the other diagonals. After a translation of the upper leg
by one lattice site the model describes two spin-$1\over2$ chains,
one antiferromagnetic and the other ferromagnetic, with 
exchange constants $\pm J$, which are coupled
via antiferromagnetic rung interaction $J'$; at $J'/J > \sqrt{2}$ the
ground state becomes a trivial product of singlets along the rungs,
a result which can be also obtained from (\ref{gensol-v0}).

\subsection{Multicritical models}
\label{subsec:multi}

The Hamiltonian couplings in the general solution (\ref{gensol-inv})
become independent of the wave function parameter $u$ if
$\lambda_{0}=\lambda_{1}^{(2,2)}=\lambda_{1}^{(1,2)}=0$, which defines
a one-parametric family of models of the form
\begin{eqnarray} 
\label{mult-inv}
&& J_{R}=2,\quad J_{L}=J_{L}'=1, \quad J_{D}=J_{D}'=(1+x)/(1-x),\nonumber\\
&& V_{RR}=V_{LL}={4\over5}{1-5x\over1-x},\quad V_{DD}={4\over5}{1+5x
\over 1-x}\,,
\end{eqnarray}
with $x\equiv (3\lambda_{1}^{(1,1)}/5\lambda_{2})\geq 0$.
The ground state degeneracy of this model is
very high: {\em any\/} wave function $\Psi_{0}(u)$ of the form
(\ref{gs}) with {\em arbitrary\/} parameter $u$ is a
ground state. One can easily calculate the overlap between two such
wave functions having different values of $u$:
\begin{equation} 
\label{overlap} 
\langle \Psi_{0}(u_{1})| \Psi_{0}(u_{2})\rangle=q^{N},\;\;
q={(1+u_{1}u_{2})^{2}\over (1+u_{1}^{2})(1+u_{2}^{2})}\leq1\,,
\end{equation} 
i.e., the two g.s.\ wave functions with different values of $u$ are
{\em asymptotically orthogonal\/} in the thermodynamic limit
$N\to\infty$ with the overlap vanishing exponentially with the
increase of $N$.  This means that the dimension of the basis of this
subspace $\{\Psi_{0}(u)\}$, i.e., the number of mutually orthogonal
ground states, is exponentially large in the thermodynamic limit.  For
$x=0$ this statement becomes trivial, because then the local
Hamiltonian projects only onto the states with the total spin of two
neighboring rungs equal to $2$, and any sequence of singlets and
triplets on the rungs which does not contain neighboring triplets is a
ground state; in this case the SU(2) symmetry of the original
Hamiltonian is spontaneously broken because the ground states are not
necessarily eigenstates of the total spin.  Later we show that at
$x={1\over5}$ the model (\ref{mult-inv}) is gapless, by constructing
the exact excitation wave function (see Sect.\ \ref{sec:BG+}).

\subsection{AKLT-type models}
\label{subsec:AKLT}

One can put $u=0$ in (\ref{gensol-inv}) and arrive at a large family
of models having the AKLT ground state formed by effective $S=1$ spins
made up of the ladder rungs. An example of such solution was presented
in our earlier paper. \cite{KM97} The simplest model within this class
can be obtained by requiring that the diagonal interactions vanish
together with the rung-rung biquadratic coupling, which is achieved by
setting $\lambda_{2}=6/5$,
$\lambda_{1}^{(1,1)}={1\over2}\lambda_{0}-{4\over5}$,
$\lambda_{1}^{(2,2)}={1\over2}\lambda_{0}+{6\over5}$,
$\lambda_{1}^{(1,2)} = 0$. The model then describes a ladder with
ferromagnetic rungs and two additional biquadratic interactions:
\begin{eqnarray}
\label{simple-AKLT} 
&& J_{R}\leq -4/5,\;\; \varepsilon_{R}=0,\quad 
J_{L}=J_{L}'=1, \quad J_{D}=J_{D}'=0,\nonumber\\
&&V_{RR}=0,\quad V_{LL}=12/5,\quad V_{DD}=-8/5\,.
\end{eqnarray}

It is worthwhile to mention that there exists a one-parametric family
of models \cite{KM97} connecting smoothly the Majumdar-Ghosh chain to
the effective AKLT chain, thus proving that the Haldane chain and the
dimerized chain are ``in the same phase'' in the sense that there is a
continuous path in the phase space which links the two states and does
not cross any singularities. However, as we will see below, this does
not generally exclude the possibility to have a phase boundary between
the Haldane (AKLT) and dimer states. A model exhibiting such a
boundary will be considered later in Sect.\ \ref{sec:BG+}.

\begin{figure}
 \mbox{\psfig{figure=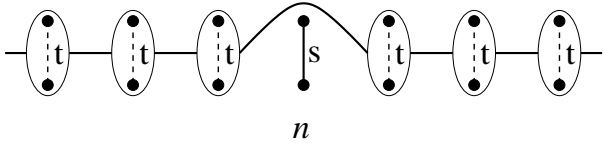,width=80mm,angle=-90}}
 \vspace{3mm}
\caption{\label{fig:aklt-exc} Schematic representation of the exact
excitation wave function (\protect\ref{aklt-singlet}) in the
``AKLT-type'' ladder model. Thick solid lines denote valence bonds
connecting effective $S=1$ spins (indicated by ovals) which are formed
by the triplet $(t)$ degrees of freedom on the ladder rungs, and $s$
denotes a singlet bond connecting two spins of the $n$-th rung.  }
\end{figure}

The AKLT-type ladder models are especially remarkable because they
admit constructing an {\em exact eigenstate of the Hamiltonian being a
singlet magnon\/} which becomes {\em gapless\/} under certain
conditions.  Let us consider the state
\begin{equation}
\label{aklt-singlet}
|n\rangle_{S}=\mbox{tr}\{g_{1}^{A}g_{2}^{A}\cdots g_{n-1}^{A} g_{n}^{S}
 g_{n+1}^{A}\cdots g_{N}^{A} \}\,,
\end{equation}
where $g_{i}^{A}= g_{i}(u=0)$ is the ground state (AKLT) matrix and
$g_{n}^{S}=g_{n}(u=\infty)=\openone \cdot |s\rangle_{n}$ is the matrix
describing a singlet bond on the $n$-th rung; a visual
VBS-representation of this state
is shown in Fig.\ \ref{fig:aklt-exc}. This state is orthogonal
to the ground state wave function
$\Psi_{0}^{\text{inv}}=\mbox{tr}(g_{1}^{A}\cdots g_{N}^{A})$. It can
be straightforwardly checked that the action of the Hamiltonian on the
state $|n\rangle_{S}$ yields nothing but the states $|n\rangle_{S}$,
$|n\pm1\rangle_{S}$:
\begin{eqnarray} 
\label{singlet-action} 
\widehat{H}|n\rangle_{S}&=&\big\{\lambda_{1}^{(1,1)}+\lambda_{1}^{(2,2)}\big\}
|n\rangle_{S} \nonumber\\
&+&{1\over2}\big\{\lambda_{1}^{(2,2)}-\lambda_{1}^{(1,1)}\big\}
\big\{ |n-1\rangle_{S}+|n+1\rangle_{S}\big\}\,.
\end{eqnarray}
[Here the ground state energy is already subtracted, i.e., the term
$-E_{0}$ is included in the local Hamiltonian $\widehat{h}$ in
(\ref{ham})]. Trivially constructing the exact  eigenstate
$|k\rangle_{S}=\sum_{n} e^{ikn}|n\rangle_{S}$
with total
momentum $k$, we obtain its energy as
\begin{equation} 
\label{singlet-disp} 
\varepsilon_{S}(k)=\lambda_{1}^{(1,1)}+\lambda_{1}^{(2,2)}+
\big\{\lambda_{1}^{(2,2)}-\lambda_{1}^{(1,1)}\big\}\cos(k)\,.
\end{equation}
The minimum of the dispersion is reached at $k=\pi$ if
$\lambda_{1}^{(1,1)}<\lambda_{1}^{(2,2)}$ and at $k=0$ if
$\lambda_{1}^{(1,1)}>\lambda_{1}^{(2,2)}$, and the excitation becomes
gapless when either $\lambda_{1}^{(1,1)}$ or $\lambda_{1}^{(2,2)}$
vanishes. Later we will see that the transition at
$\lambda_{1}^{(1,1)}=0$ is into a spontaneously dimerized phase, and
the transition at $\lambda_{1}^{(2,2)}=0$ is specific for the
AKLT-type models.

Far from the phase boundaries one should expect that the singlet
magnon branch will be high in energy, and the lowest excitation will
be a usual triplet magnon being a soliton in the string
order. \cite{sol-string,NM96} 

\subsection{Rung-dimer models with exact triplet excitations}
\label{subsec:RDtriplet}

Inside the general family (\ref{gensol-v0}), (\ref{ineq-v0}),
describing models with rung-dimer ground state, it is
worthwhile to consider a special subclass of models satisfying the
additional condition
\begin{equation} 
\label{cond-RD-triplet} 
J_{L}^{(-)}+J_{D}^{(-)}=0\,.
\end{equation}
Those models are remarkable because it is a straightforward exercise
to check that the simplest excitation resulting from promoting
one ladder rung from singlet to a triplet,
\begin{eqnarray} 
\label{rd-triplet} 
&&|k\rangle_{t}=\sum_{n} e^{ikn}|n\rangle_{t}\,\\
&&|n\rangle_{t}=|s\rangle_{1}|s\rangle_{2}\cdots
|s\rangle_{n-1}|t\rangle_{n}|s\rangle_{n+1}\cdots |s\rangle_{N},\nonumber
\end{eqnarray}
is an exact eigenfunction with the energy
\begin{eqnarray} 
\label{rd-t-disp} 
\varepsilon_{t}(k)&=&J_{R}-{1\over2}(V_{LL}+V_{DD}+3V_{RR})\nonumber\\
	&+&2(J_{L}^{(+)}-J_{D}^{(+)})\cos k\,.
\end{eqnarray}
This exact excitation is of course not always the lowest one [e.g.,
for the model of $\Delta$-chain, $x=1-y$ in (\ref{ss-delta}), it is
dispersionless and has a high energy equal to $1$], but in some cases
the dispersion law (\ref{rd-t-disp}) has gapless or almost gapless
points and thus the triplet excitation defined above is relevant at
least in some interval of momenta $k$. One interesting realization is the
ladder model with leg-leg biquadratic coupling, given by
\begin{equation} 
\label{rd-mik} 
J_{L}=J_{L}'={1\over4}V_{LL}, \quad J_{R}\geq 4 J_{L}\geq 0\,,
\end{equation}
with all the other couplings being zero. The dispersion
(\ref{rd-t-disp}) in this case reads as
$\varepsilon_{t}(k)=J_{R}-2J_{L}(1-\cos k)$, and becomes gapless  at
$k=\pi$ for $J_{L}\to
{1\over4}J_{R}$. Another interesting model within this class will be
considered later in Sect.\ \ref{sec:BG+}.

\section{Spontaneously dimerized models: phase boundaries and
elementary excitations}
\label{sec:dim}

For the dimerized matrix product ansatz $\Psi_{0}^{\text{dim}}$ the
local ground states entering  two matrix products $g_{1}(u_{1})\cdot
g_{2}(u_{2})$ and $g_{1}(u_{2})\cdot g_{2}(u_{1})$ are
\begin{eqnarray} 
\label{lgs-dim} 
&&|\psi_{00}^{(g)}\rangle=\big[3+(u_{1}u_{2})^{2}\big]^{-1/2}
\big\{u_{1}u_{2}|ss\rangle
-\sqrt{3}|tt\rangle_{00}\big\},\nonumber\\
&&|\psi_{1\mu}^{(g,1)}\rangle=[1+f^{2}/2]^{-1/2}\big\{ 
(f/2)(|st_{\mu}\rangle+|t_{\mu}s\rangle) -|tt\rangle_{1\mu}\big\} \nonumber\\
&&|\psi_{1\mu}^{(g,2)}\rangle=\mbox{sgn}(u_{1}-u_{2})
{1\over\sqrt{2}}(|st_{\mu}\rangle-|t_{\mu}s\rangle),
\end{eqnarray}
where $f\equiv (u_{1}+u_{2})/\sqrt{2}$, and the remaining multiplets
can be chosen as
\begin{eqnarray} 
\label{les-dim} 
&&|\psi_{00}^{(e)}\rangle=\big[3+(u_{1}u_{2})^{2}\big]^{-1/2}
\big\{\sqrt{3}|ss\rangle
+u_{1}u_{2}\,|tt\rangle_{00}\big\},\nonumber\\
&&|\psi_{1\mu}^{(e)}\rangle=[2+f^{2}]^{-1/2}\big\{ 
f|tt\rangle_{1\mu}+|st_{\mu}\rangle+|t_{\mu}s\rangle\big\} \nonumber \\
&&|\psi_{2\mu}^{(e)}\rangle=|tt\rangle_{2\mu}.
\end{eqnarray}
Note that at $u_{1}=u_{2}$ the multiplet $|\psi_{1\mu}^{(g,2)}\rangle$
disappears from the set of local ground states and becomes an
eigenstate, in accordance with Eqs.\ (\ref{lgs-inv}),
(\ref{les-inv}). Performing the same procedure as for undimerized
case, we arrive at the following general family of solutions:
\begin{eqnarray}
&& J_{R}={1\over6}\widetilde{A}\widetilde{B}\lambda_{0}
+{1\over2}\widetilde{C}\lambda_{1} +{5\over6}\lambda_{2},\quad
\varepsilon_{R}=0,\nonumber\\ 
&&J_{L}=J_{L}'=-{1\over12}\widetilde{A}(\widetilde{A}+\widetilde{B})\lambda_{0}
-{1\over4} \widetilde{C} \lambda_{1} +{5\over12}\lambda_{2},\nonumber\\
&&J_{D}^{(+)}=-{1\over12}\widetilde{B}(\widetilde{A}+\widetilde{B})
\lambda_{0} -{1\over4}\lambda_{1}+{5\over12}\lambda_{2},\nonumber\\
&& J_{D}^{(-)}=-{2(u_{1}+u_{2})\over
(u_{1}+u_{2})^{2}+4}\lambda_{1},\nonumber\\ 
&& V_{RR}=-{1\over3}\widetilde{A}\widetilde{B}\lambda_{0}
+\widetilde{C}\lambda_{1}+{1\over3}\lambda_{2},\nonumber\\
&& V_{LL}={1\over3}\widetilde{A}(\widetilde{A}+\widetilde{B})\lambda_{0}
-\widetilde{C}\lambda_{1} +{1\over3}\lambda_{2},\nonumber\\
&&V_{DD}={1\over3}\widetilde{B}
(\widetilde{A}+\widetilde{B})\lambda_{0}-\lambda_{1}+{1\over3} \lambda_{2},
\label{gensol-dim} 
\end{eqnarray}
with the energy density per rung
\begin{equation} 
\label{e0-dim} 
E_{0}=-{1\over16}\lambda_{0}-{3\over16}\lambda_{1}-{5\over16}\lambda_{2}\,.
\end{equation}
Here the quantities $\lambda_{j}\geq0$, $j=0,1,2$ denote the local
eigenvalues corresponding to the multiplets $|\psi_{jm}^{(e)}\rangle$, 
and the
factors $\widetilde{A}$, $\widetilde{B}$, $\widetilde{C}$ are defined
as follows:
\begin{eqnarray} 
\label{not-dim} 
&&\widetilde{A}={u_{1}u_{2}+3\over (u_{1}^{2}u_{2}^{2}+3)^{1/2}},\quad 
\widetilde{B}={u_{1}u_{2}-3\over (u_{1}^{2}u_{2}^{2}+3)^{1/2}},\nonumber\\
&&\widetilde{C}={(u_{1}+u_{2})^{2}-4 \over(u_{1}+u_{2})^{2}+4}
\end{eqnarray}
Now we proceed to the study of general properties of the dimerized
models derived above.

\subsection{Phase boundaries}
\label{subsec:bound}

Comparing the general solutions (\ref{gensol-inv}) and
(\ref{gensol-dim}), one can easily see that they match
each other if
\begin{equation} 
\label{match} 
u_{1}=u_{2}=u,\quad \lambda_{1}^{(1,1)}=\lambda_{1}^{(1,2)}=0,
\end{equation}
then $\lambda_{1}^{(2,2)}$, $\lambda_{0}$, $\lambda_{2}$ in
(\ref{gensol-inv}) should be identified respectively
with $\lambda_{1}$, $\lambda_{0}$, $\lambda_{2}$ in (\ref{gensol-dim}).
 The original Hamiltonian (\ref{ham}) is
translational invariant, thus the condition (\ref{match}) defines a
{\em critical line\/} in the space of the model parameters where the
second-order quantum phase transition into spontaneously dimerized
phase occurs. 

It should be mentioned that such a transition for ladders was first
studied by Nersesyan and Tsvelik \cite{NersesyanTsvelik97} within a
field-theoretical approach in the approximation of weak inter-leg
coupling. They found that leg-leg biquadratic exchange is a relevant
perturbation, so that at sufficiently strong $V_{LL}$ the ladder can
enter into a spontaneously dimerized {\em non-Haldane spin liquid
phase,\/} with the elementary excitations being soliton-antisoliton
pairs. Later we have proposed explicit examples of solvable models
exhibiting such properties,\cite{KM98prl} and from the general
solution (\ref{gensol-dim}) it follows that other interactions may be
responsible for this transition as well. (For example, it is possible
to get a spontaneously dimerized ladder with $V_{LL}=0$). However, one
can see that in MP-solvable ladder models described by (\ref{match})
the dimerization transition has certain peculiarities: It is easy to
calculate spin-spin and dimer-dimer correlation functions
$C_{S}(n)=\langle S^{z}_{1,i} S^{z}_{1,i+n}\rangle$ and
$C_{D}(n)=\langle D_{i}D_{i+n}\rangle$,  $D_{i}={\mathbf
S}_{1,i}\cdot({\mathbf S}_{1,i+1}-{\mathbf S}_{1,i-1})$ being the
dimerization order parameter:
\begin{eqnarray} 
\label{correl} 
C_{S}(n)&=&(u_{2}^{2}+3)^{-1}
(Z_{12}Z_{21})^{n} 
(\delta_{n,2k}+Z_{21}\delta_{n,2k+1}), \\
C_{D}(n)&=&{36(u_{1}-u_{2})^{2}\over(u_{1}^{2}+3)^{2}(u_{2}^{2}+3)^{2}},\quad 
Z_{ab}={(u_{a}-1)(u_{b}+1)\over 3+u_{b}^{2}}
\,,\nonumber
\end{eqnarray}
here $a,b=1,2$. One can see that the dimer correlations exhibit
long-range order vanishing for $u_{1}\to u_{2}$ or $u_{1,2}\to\infty$,
but remarkably there is no exponential tail present in the 
correlations.  Spin-spin correlations always decay
exponentially, and the correlation length $\xi=-1/\ln(Z_{12}Z_{21})$
does not show any singular behavior at $u_{1}\to u_{2}$, but diverges
at $u_{1,2}\to\infty$; however, there is no long-range spin order at
$u_{1,2}\to\infty$ since the amplitude of correlations in this limit
vanishes. From the field-theoretical point of view, such peculiarities
correspond to the ``fine tuning'' of the theory parameters which
makes certain pre-exponential factors zero.

It is easy to check that the general family of MP-solvable
spontaneously dimerized models (\ref{gensol-dim}) includes one 
model of the multicritical type:\cite{KM98prl} if $u_{1}=-u_{2}=u$ and
$\lambda_{0}=0$, wave functions $\Psi_{0}^{\text{dim}}(u)$ with any
$u$ have the same energy and are degenerate ground states. This
solution can also be matched to the translational invariant family
(\ref{gensol-inv}) at $\lambda_{1}^{(1,1)}=\lambda_{0}=0$, $u=0$.

\subsection{Elementary excitations in dimerized phase}

In order to discuss the structure of elementary excitations in the
dimerized phase, let us consider a model\cite{KM98prl} defined by setting 
\begin{eqnarray*} 
&& u_{1}=-u_{2}=u,\quad \lambda_{0}(u^{4}+10u^{2}+5)=16/3,\\
&&\lambda_{1}=\lambda_{0}(3u^{4}\!+ 14u^{2}\!+15)/8,\;
\lambda_{2}=\lambda_{0}(5u^{4}\!+18u^{2}\!+9)/8,
\end{eqnarray*}
which yields the Hamiltonian (\ref{ham}) with the following
couplings:
\begin{eqnarray} 
\label{familyC} 
&&V_{DD}=-V_{RR}=J_{R}={8\over3}{(u^{2}-1)(u^{2}+3)\over u^{4}+10u^{2}+5},\\
&& V_{LL}={4\over3}{5u^{4}+2u^{2}+9\over  u^{4}+10u^{2}+5},\;\;
 J_{L}=J_{L}'=1,\;\; J_{D}=J_{D}'=0.\nonumber
\end{eqnarray}
The ground state energy per rung given by
\[
E_{0}=-{7u^{4}+22u^{2}+19 \over 4(u^{4}+10u^{2}+5)}\,.
\]

\begin{figure}
 \mbox{\psfig{figure=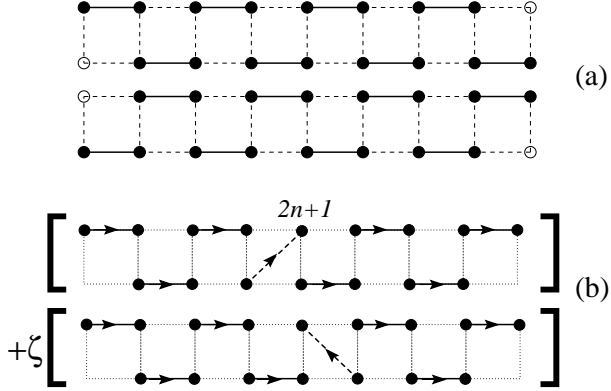,width=80mm,angle=-90}}
 \vspace{3mm}
\caption{\label{fig:ntgsexc} (a) Two degenerate ground states of the
``generic'' dimerized model (\protect\ref{familyC}) with $u=1$;  (b)
the soliton states $|n\rangle_{t,s}^{\zeta}$
used in Eq.\ (\protect\ref{genexc}), in the same special case 
$u=1$.  Thick
solid lines indicate singlet bonds, and thick dashed lines can be
either singlets or triplets. Arrows indicate the ``direction'' of the
singlet bonds [i.e.,
$|s_{1\rightarrow2}\rangle=2^{-1/2}(|\uparrow_{1}\downarrow_{2}\rangle
-|\downarrow_{1}\uparrow_{2}\rangle)$].}
\end{figure}

A ``generic'' example from this family is the model at $u=\pm1$, when
one gets purely biquadratic interchain interaction. It is remarkable
first of all because of simplicity of its ground state, which is just
a checkerboard-type product of singlet bonds along the ladder legs
(see Fig.\ \ref{fig:ntgsexc}a); second, this model lies within the class
of Hamiltonians considered by Nersesyan and
Tsvelik.\cite{NersesyanTsvelik97}

Elementary excitations of the model (\ref{familyC}) are pairs of
solitons in the dimer order, and the scattering states of can be
studied with the help of the following variational MP ansatz for a
single soliton: \cite{KM98prl}
\begin{eqnarray} 
\label{genexc}
&&|p\rangle_{t,s}^{\zeta}=\sum_{n} 
e^{ip(2n+1)}|n\rangle_{t,s}^{\zeta} \,,\\
&&|n\rangle_{s,t}^{\zeta}=\prod_{i=1}^{n}
\{g_{2i-1}(-u) g_{2i}(u)\}  g_{2n+1}^{s,t}\!\!
\prod_{i=n+1}^{N} \!\! g_{2i}(-u) g_{2i+1}(u)
,\nonumber\\
&&g^{s}_{\zeta}= g(u) -\zeta g(-u),\quad
g^{t}_{\zeta,\mu}= \sigma^{\mu} g(u) +\zeta
g(-u)\sigma^{\mu}\,. \nonumber
\end{eqnarray}
Here we for the moment assume that the ladder has $2N+1$ rungs and
periodic boundary conditions, so that the one-soliton wave function is
well defined, $\mu=0,\pm1$ denotes the $z$-projection of spin of the
triplet excitation, and $\zeta=\pm1$ is the parity of a single
soliton. The momenta are defined in terms of the Brillouin zone of
non-dimerized ladder, so that $p\in[0,\pi]$.
In case of the ``generic'' model with $u=\pm1$ the states
$|n\rangle_{t,s}^{\zeta}$ can be visualized as singlet or triplet
diagonal bonds separating two ``checkerboard-dimer'' ground states, as
shown in Fig.\ \ref{fig:ntgsexc}b.

The lowest energy for the variational excitation (\ref{genexc}) is
always reached for the odd-parity state ($\zeta=-1$), and the
variational gap for the Haldane triplet is always higher than the gap
for a soliton-antisoliton pair.\cite{KM98prl} The energy of the
elementary excitation (soliton pair)  for the scattering states
is given by
\[
\widetilde{E}(k,q)=\varepsilon_{s,t}\big[ (k+q)/2\big]
+\varepsilon_{s,t}\big[ (k-q)/2\big]\,,
\] 
where $k$ and $q$ are the total and relative momentum.  

The following expressions for the variational 
gaps can be obtained:
\begin{equation} 
\label{ntgaps} 
\Delta_{ss}^{(-)}={4u^{4}\over(u^{2}+3)^{2}}\,,\quad
 \Delta_{tt}^{(-)}={4\over(u^{2}+3)^{2}}\,.
\end{equation}
At $u\to0$ the odd-singlet soliton gap goes to zero, indicating the
second-order transition to the Haldane phase, and at $u\to\infty$ the
odd-triplet soliton gap vanishes, signaling another second-order
transition into the rung-dimer phase.  Thus, transition to the Haldane
(AKLT) phase at $u=0$ is governed by a vanishing singlet-singlet gap,
and the other transition to the rung-dimer phase at $u\to\infty$ is
determined by the closing singlet-triplet gap; later in Sect.\
\ref{sec:BG+} we will see that the same property holds also for a
different model, which suggests that it has a general character.

\section{Generalized Bose-Gayen model}
\label{sec:BG+}

In this section we introduce a toy model which allows one to view
simultaneously 
nearly all possibilities offered by the MP approach. Let us consider
the model described by the Hamiltonian (\ref{ham}) with
\begin{equation} 
\label{ham-bg+} 
J_{L}=J_{L}'=1,\; J_{R}=y_{1},\; J_{D}=J_{D}'=y_{2},\; V_{RR}=0.
\end{equation}
At $y_{2}=1$ and $V_{LL}=V_{DD}=0$ this model was first introduced by
Bose and Gayen, \cite{BoseGayen93+} and its generalization to the case
of arbitrary $y_{2}$ was recently considered by  Weihong, Kotov and
Oitmaa.\cite{Weihong+} 

The line $y_{2}=1$ is quite peculiar because, as was shown by Xian,
\cite{BoseGayen93+} in this case the operator of the rung interaction
commutes with the rest of the Hamiltonian, which enables one to
classify the eigenstates by the total spin of each rung. At
$y_{1}>\varepsilon_{0}\approx 1.4$ the exact ground state is just a
product of singlet bonds along the rungs, and at
$y_{1}<\varepsilon_{0}$ the ground state coincides with that of the
effective $S=1$ Haldane chain whose $S=1$ spins are formed by the
pairs of $S={1\over2}$ spins on the ladder rungs; $\varepsilon_{0}$ is
exactly the ground state energy per spin of the $S=1$ Haldane chain, and at
$y_{1}=\varepsilon_{0}$ a first-order phase transition from the
Haldane phase to the rung-dimer phase occurs. \cite{BoseGayen93+} When
$y_{2}\not=1$, this point of transition develops into a line as shown
numerically by Weihong et al.\cite{Weihong+} It is also worthwhile to
mention that when the rung exchange $y_{1}$ is allowed to alternate,
the system exhibits a nice sequence of first-order phase
transitions. \cite{NiggUimZitt97}

It is natural to try to ``deform'' the model by including biquadratic
couplings $V_{LL}$ and $V_{DD}$ in such a way that the Haldane phase
becomes simply the AKLT phase, and then its ground state can be
expressed through a matrix product wave function. It is a
straightforward exercise to check that the general family
(\ref{gensol-inv}) of MP-solvable models with translational invariant
ground state reduces to the form (\ref{ham-bg+}) under the following
choice of parameters:
\begin{eqnarray} 
\label{bg-cond-aklt}
&&\lambda_{1}^{(1,1)}=-{1\over2}y_{1}+{2\over5}(4y_{2}-1) 
\geq 0,\quad \lambda_{1}^{(1,2)}=0,\nonumber\\
&&\lambda_{1}^{(2,2)}=-{1\over2}y_{1}+{2\over5}(4-y_{2}) 
\geq 0,\quad u=0,\\
&&\lambda_{0}=-y_{1}+{4\over5}(1+y_{2})\geq 0,\quad
\lambda_{2}={6\over5}(1+y_{2})\geq 0\,.\nonumber
\end{eqnarray}
Then the biquadratic couplings are expressed through $y_{2}$ as
\begin{equation} 
\label{bg-mp} 
V_{LL}={4\over5}(3-2y_{2}),\quad V_{DD}={4\over5}(3y_{2}-2)\,.
\end{equation}
This model has the AKLT-type ground state with $u=0$ under the
conditions imposed by the inequalities (\ref{bg-cond-aklt}), with the
ground state energy per rung
\begin{equation} 
\label{e0-bg-aklt} 
E_{0}^{\text{AKLT}}={1\over4} y_{1} -{13\over20}(1+y_{2})\,.
\end{equation}

From the general analysis presented in previous sections we know
that vanishing $\lambda_{0}$ corresponds to the phase transition into
the rung-dimer phase, $\lambda_{2}=0$ determines the transition into the
completely polarized ferromagnetic phase, and $\lambda_{1}^{(1,1)}=0$
gives the phase boundary of a transition into a spontaneously dimerized
state, only the last transition being of the second order. 
The Hamiltonian (\ref{ham-bg+}), (\ref{bg-mp}) has a special symmetry
(cf.\ Ref.\ \onlinecite{Weihong+}):
exchanging the two spins of every second rung is equivalent to
exchanging the leg coupling $1$ with the diagonal coupling $y_{2}$
with simultaneous exchanging $V_{LL}$ and $V_{DD}$, so
that there is a one-to-one correspondence between the eigenstates of
the Hamiltonian at $y_{2}>1$ and $0<y_{2}<1$, defined as
\begin{equation} 
\label{bg-symmetry} 
y_{1}\mapsto y_{1}/y_{2},\quad y_{2}\mapsto 1/y_{2},\quad y_{2}\geq 0\,.
\end{equation}
Exploiting this symmetry, one can obtain the ``mirror'' of the
second-order transition line $\lambda_{1}^{(1,1)}=0$ in the other
half-plane; this happens to be exactly the line defined by the
equation $\lambda_{1}^{(2,2)}=0$.  The line $\lambda_{0}=0$ is
invariant under the transformation (\ref{bg-symmetry}).  At the
intersection of lines $\lambda_{0}=0$ and $\lambda_{1}^{(2,2)}=0$ we
have the multicritical point $(y_{1}=2,y_{2}={3\over2})$ of the type
$\ldots g(u)\, g(u)\ldots$ corresponding to the model (\ref{mult-inv})
with $x={1\over5}$ (see Sect.\ \ref{subsec:multi}), and its ``mirror''
is obviously another multicritical point
$(y_{1}={4\over3},y_{2}={2\over3})$ of the type $\ldots g(u)\, g(-u)\,
g(u)\, g(-u)\ldots$ (see Ref.\ \onlinecite{KM98prl} and Sect.\
\ref{subsec:bound}), lying at the intersection of the lines
$\lambda_{1}^{(1,1)}=0$ and $\lambda_{0}=0$.

Since the model (\ref{ham-bg+}), (\ref{bg-mp}) 
belongs to the class of AKLT-type models considered in
Sect.\ \ref{sec:trans-inv}, it is possible to write the exact
wave function of the singlet magnon in the AKLT phase whose softening
drives the second-order phase transitions at
$y_{2}={1\over4}+{5\over16}y_{1}$ and
$y_{2}=4-{5\over4}y_{1}$. According to (\ref{singlet-disp}), its
dispersion law reads as
\begin{equation} 
\label{bg-singlet} 
\varepsilon(k)=-y_{1}+{6\over5}(1+y_{2}) +2(1-y_{2})\cos(k)\,,
\end{equation}
so that the excitation becomes dispersionless at  $y_{2}=1$. Near the
lines $\lambda_{1}^{(1,1)}=0$, $\lambda_{1}^{(2,2)}=0$
those singlet excitation should be the lowest ones, and
their condensation  determines the physics of 
second-order phase transition.  On the
line of transition into the dimerized phase,
$y_{2}={1\over4}+{5\over16}y_{1}$, the gap closes at $k=\pi$, and on
the other line $y_{2}=4-{5\over4}y_{1}$ the gap closes at
$k=0$. 

The  Hamiltonian (\ref{ham-bg+}), (\ref{bg-mp}) satisfies also 
the special $u=\infty$ solution (\ref{gensol-v0}), which means that it
has a rung-dimer ground state within the region determined by the
conditions (\ref{ineq-v0}), (\ref{3x3v0}); those conditions take the
form
\begin{eqnarray} 
\label{bg-cond-dimer} 
&&\mu_{0}=\widetilde{\mu}_{1}^{(3)}=y_{1}-{4\over5}(1+y_{2})\geq
0,\nonumber\\
&&\widetilde{\mu}_{1}^{(1)}={1\over2}y_{1}+{2\over5}(2y_{2}-3)\geq
0,\nonumber\\
&&\widetilde{\mu}_{1}^{(2)}={1\over2}y_{1}+{2\over5}(2-3y_{2})\geq
0,\\
&&\mu_{2}=y_{1}+{2\over5}(1+y_{2})\geq 0\,,\nonumber
\end{eqnarray}
and the ground state energy per rung for the rung-dimer phase is
\begin{equation} 
\label{e0-bg-dim} 
E_{0}^{\text{RD}}=-{3\over4}y_{1}+{3\over20}(1+y_{2})\,.
\end{equation}
It should be remarked that the equations (\ref{bg-cond-dimer})
indicate only the region of {\em guaranteed\/}
stability of the rung-dimer phase,  and
generally they {\em do not\/} necessarily correspond to the real phase
boundaries (recall that the conditions of stability imposed
by the inequalities (\ref{lambdaeqs}) are {\em sufficient\/} but not
{\em necessary\/});
for example, a naive application of the Eqs.\
(\ref{bg-cond-dimer}) to the original Bose-Gayen model would give
$y_{1c}=2$ as the boundary of the rung-dimer phase, while the correct
region of stability, according to Xian, \cite{BoseGayen93+} is wider,
$y_{1c}\approx 1.4$. 
However, in the present case the lines $\mu_{0}=0$,
$\widetilde{\mu}_{1}^{(1)}=0$, $\widetilde{\mu}_{1}^{(2)}=0$ are {\em
true phase boundaries.\/} The line $\mu_{0}=0$ obviously
determines the first-order phase transition because it precisely
coincides with the $\lambda_{0}=0$ line, so that we know exact ground
states on both sides.  Further, one can show that the lines
$\widetilde{\mu}_{1}^{(1)}=0$, $\widetilde{\mu}_{1}^{(2)}=0$ determine
second-order transitions. Indeed,
the model (\ref{ham-bg+}), (\ref{bg-mp})
satisfies the condition (\ref{cond-RD-triplet}), and thus the
wave function (\ref{rd-triplet}) is an 
exact excited eigenstate, with the dispersion given by
\begin{equation} 
\label{triplet-disp} 
\varepsilon_{t}(k)=y_{1}-{2\over5}(1+y_{2})+2(1-y_{2})\cos k\,.
\end{equation}
It is easy to see that the $k=\pi$ gap closes at the
$\widetilde{\mu}_{1}^{(1)}=0$ line, and the $k=0$ gap closes at the
$\widetilde{\mu}_{1}^{(2)}=0$ line.  Note that the symmetry transformation
(\ref{bg-symmetry}) leaves the dispersions (\ref{bg-singlet}) and
(\ref{triplet-disp}) invariant, except for the change $k\mapsto
\pi-k$.

Boundaries of the ferromagnetic phase can be obtained by requiring that
$\lambda_{2}$ is the lowest eigenvalue of the local Hamiltonian (in
this case only relative sign of $\lambda$'s matters), which gives 
\[
{5\over4}y_{1}+4 < y_{2} < -1\,.
\]
It is interesting to remark that the same conditions of stability of
the FM phase can be obtained from the standard lowest order spin-wave
theory: one has two branches of ferromagnons with the dispersion laws
\begin{eqnarray*}
\varepsilon^{(+)}(k)&=&-{6\over5}(1+y_{2})(1-\cos k),\\
\varepsilon^{(-)}(k)&=&-{6\over5}(1+y_{2})-y_{1}+2(1-y_{2})\cos k\,,
\end{eqnarray*}
and demanding the excitation energies to be positive one arrives at
the same result. The transition at the line $y_{2}=-1$ is
of the first order (the entire branch $\varepsilon^{(+)}(k)$ of
``usual'' ferromagnons collapses), while for the line
$y_{2}={5\over4}y_{1}+4$ the situation is different: the $k=\pi$ gap
of the ``optical'' ferromagnons branch $\varepsilon^{(-)}(k)$ closes,
which suggests a second-order transition.

\begin{figure}
 \mbox{\psfig{figure=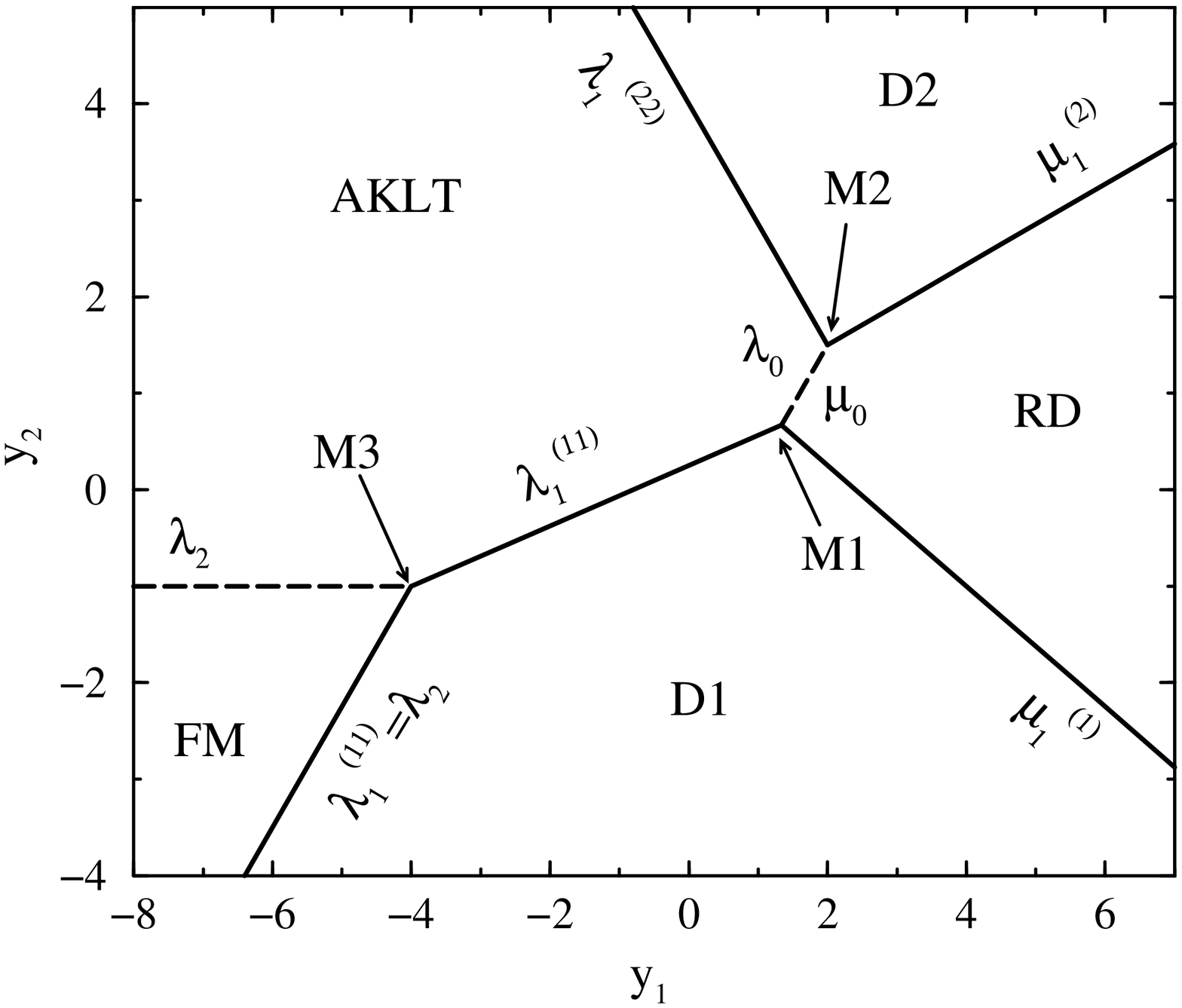,width=80mm,angle=-90}}
 \vspace{3mm}
\caption{\label{fig:bg-diag} Phase diagram of the generalized
Bose-Gayen model (\protect\ref{ham-bg+}), (\protect\ref{bg-mp}). AKLT
and FM denote the effective Affleck-Kennedy-Lieb-Tasaki ($u=0$) and
ferromagnetic phase, respectively, RD stays for the rung-dimer phase,
and D1, D2 indicate phases with spontaneously broken symmetry where
exact ground state is not known. The transitions AKLT-RD, AKLT-FM
(shown with thick dashed lines) are of the first order, while all the
other ones (shown with thick solid lines) are second-order
transitions.  Points M1 $({4\over3},{2\over3})$ and M2 $(2,{3\over2})$
are multicritical with exponential degeneracy of the ground state, and
M3 $(-4,-1)$ is also a multicritical point but of different nature.
The indicated eigenvalues of the local Hamiltonian vanish on the
corresponding transition lines (on the FM-D1 line
$\lambda_{2}=\lambda_{1}^{(1,1)}$).  }
\end{figure}

Gathering up all that, one arrives at the complete phase diagram 
presented in Fig.\ \ref{fig:bg-diag}.  It is
interesting that the model (\ref{ham-bg+}), (\ref{bg-mp}), despite its
simplicity, has a rich phase diagram with five phases and three
multicritical points.  All transition lines are exact, and only inside
the phases marked D1 and D2 the ground state is not known exactly.
Existence of second-order transition boundaries to the AKLT and RD
phases means that the symmetry of D1, D2 is lower than that of the
AKLT phase; this spontaneously broken symmetry can be only a discrete
symmetry connected with the spatial parity of the system, which means
that the ground state in D1, D2 is twofold degenerate. The AKLT-D1
boundary matches our condition (\ref{match}) for the phase transition
into the dimerized state with spontaneously broken translational
symmetry, which indicates that D1 is dimerized. The properties of D2
are equivalent to those of D1 in the sense that the two phases are
connected by the symmetry transformation (\ref{bg-symmetry}).

\section{Summary}
\label{sec:summary}

We have studied a class of generalized $S={1\over2}$ ladder models
admitting exact solution for the ground state in terms of finitely
correlated, or matrix product (MP) states.  We use two different MP
ans\"atze for the ground state wave function and respectively obtain
two families of Hamiltonians with exact ground states, one being
translationally invariant and the other one spontaneously dimerized.

The two families have non-empty intersection, which enables us to
determine the boundary of a second-order phase transition from
translationally invariant to dimerized phase.  We show that the
behavior of the dimer order parameter and spin correlation functions
at the transition is peculiar: The spin-spin correlation length $\xi$
can be either finite or diverge when approaching the transition point,
but in the latter case there is no long-range spin order since the
preexponential factor vanishes {\em exactly\/} at the transition; the
dimer order correlation function is just a distance-independent
constant term vanishing at the transition.

We show that for a certain class of translationally invariant models
(having effectively a spin-$1$ AKLT ground state formed by the triplet
degrees of freedom at each rung) it is possible to write an exact wave
function for the singlet branch of elementary excitations whose gap
closes exactly at two second-order phase transition boundaries; one of
those boundaries corresponds to the transition into dimerized phase,
and the origin of the other one is different.

For excitations in the spontaneously dimerized phase we propose a
simple variational MP-type ansatz describing the elementary excitation
as a pair of solitons in dimer order; we show that variational
estimate for the gap goes to zero at the phase boundary.

As an illustration of the technique and the ideas of the present approach,
we also propose a toy model being an MP-solvable deformation of the
``composite spin'' model recently considered by Bose and Gayen
\cite{BoseGayen93+} and later by Weihong {\em et al.\/}
\cite{Weihong+} One can obtain exactly its complete phase diagram,
including several lines of phase transitions of the first and second
order.

\acknowledgements

We would like to thank S. Brehmer for many fruitful conversations and
A. M. Tsvelik for the discussion of the results.  A.K.\ gratefully
acknowledges the hospitality of Hannover Institute for Theoretical
Physics.  This work was supported in part by the German Ministry for
Research and Technology (BMBF) under the contract 03MI4HAN8 and by the
Ukrainian Ministry of Science (grant 2.4/27).

\newpage

\newpage

\newpage

\newpage

\begin{references}

\bibitem[\ast]{perm} Permanent address: Institute of Magnetism,
36(b) Vernadskii av., 252142 Kiev, Ukraine, email address:
kolezhuk@joanna.ru.kiev.ua 

\bibitem[\dag]{email} email address: mikeska@itp.uni-hannover.de 

\bibitem{DagottoRice96}  See for a review
E. Dagotto and T. M. Rice, Science {\bf 271}, 618 (1996).

\bibitem{BoseGayen93+} I. Bose and S. Gayen, Phys. Rev. B {\bf 48},
10653 (1993); see also Y. Xian, Phys. Rev. B {\bf 52}, 12485 (1995);

\bibitem{BMN96} S. Brehmer, H.-J. Mikeska and U. Neugebauer,
J. Phys: Condens. Matter {\bf 8}, 7161 (1996).

\bibitem{BKMN98} S. Brehmer, A. K. Kolezhuk, H.-J. Mikeska and
U. Neugebauer, J. Phys.: Condens. Matter {\bf 10}, 1103 (1998).

\bibitem{KM97} A. K. Kolezhuk and H.-J. Mikeska, Phys. Rev. B {\bf
56}, R11380 (1997).

\bibitem{NersesyanTsvelik97} A. A. Nersesyan and A. M. Tsvelik,
Phys. Rev. Lett. {\bf 78}, 3939 (1997).

\bibitem{KM98prl} A. K. Kolezhuk and H.-J. Mikeska, 
Phys. Rev. Lett. {\bf 80}, 2709 (1998).

\bibitem{Weihong+} Z. Weihong, V. Kotov, and J. Oitmaa, e-print
cond-mat/9711006.

\bibitem{Fannes+} M. Fannes, B. Nachtergaele and R. F. Werner,
Europhys. Lett. {\bf 10}, 633 (1989); Commun. Math. Phys. {\bf 144}, 443
(1992).

\bibitem{Klumper+} A. Kl\"umper, A. Schadschneider and
J. Zittartz, J. Phys.  A {\bf 24}, L955 (1991); Z. Phys. B {\bf 87},
281 (1992); Europhys. Lett. {\bf 24}, 293 (1993).

\bibitem{AKLT} I. Affleck, T. Kennedy, E. H. Lieb and H. Tasaki, Phys.
Rev. Lett. {\bf 59}, 799 (1987);
Commun. Math. Phys. {\bf 115}, 477 (1988).

\bibitem{KMY97} A. K. Kolezhuk, H.-J. Mikeska and Shoji Yamamoto,
Phys. Rev. B {\bf 55}, R3336 (1997).

\bibitem{NM96} U. Neugebauer and H.-J. Mikeska, Z. Phys. B {\bf 99},
151 (1996). 

\bibitem{TotsukaSuzuki95} K. Totsuka and M. Suzuki, J. Phys.:
Condens. Matter {\bf 7}, 1639 (1995).

\bibitem{OstlundRommer95} S. Ostlund and S. Rommer,
Phys. Rev. Lett. {\bf 75}, 3537 (1995).

\bibitem{MajumdarGhosh69} C. K. Majumdar and D. K. Ghosh, J. Math. Phys. {\bf
10}, 1399 (1969).

\bibitem{NiggZitt96} H. Niggemann and J. Zittartz,
Z. Phys. B {\bf 101}, 289 (1996).

\bibitem{SS81} B. S. Shastry and B. Sutherland, Phys. Rev. Lett.
{\bf 47}, 964 (1981).

\bibitem{delta-chain} F. Monti and A. S\"ut\"o, Phys. Lett. {\bf
156A}, 197 (1991); T. Nakamura and K. Kubo, Phys. Rev. B {\bf 53},
6393 (1996); D. Sen, B. S. Shastry, R. E. Walstedt, and R. Cava,
Phys. Rev. B {\bf 53}, 6401 (1996).

\bibitem{Dmitriev+97} D. V. Dmitriev, V. Ya. Krivnov, and
A. A. Ovchinnikov, Z. Phys. B {\bf 103}, 193 (1997).

\bibitem{sol-string} S. Knabe, J. Stat. Phys. {\bf 52}, 627 (1988);
G. F\'ath and J. S\'olyom, J. Phys.: Condens. Matter {\bf 5}, 8983
(1993).

\bibitem{NiggUimZitt97} H. Niggemann, G. Uimin, and J. Zittartz,
J. Phys.: Condens. Matter {\bf 9}, 9031 (1997).


\end{references}
\end{document}